\documentclass[aps, prb, reprint, amsmath, amssymb, superscriptaddress, floatfix]{revtex4-2}
\usepackage{graphicx}
\usepackage{bm}			% Bold math
\usepackage{tabularx}
\usepackage[colorlinks, citecolor=blue, linkcolor=blue]{hyperref}

\begin{document}
\title{Time-resolved observation of magnon splitting into vortex gyration and Floquet spin waves}
\author{T. Devolder} % \orcidlink{0000-0001-7998-0993}}
\email{thibaut.devolder@cnrs.fr}
\author{R. Lopes Seeger}
\affiliation{Universit\'e Paris-Saclay, CNRS, Centre de Nanosciences et de Nanotechnologies, Palaiseau, France}
\author{C. Heins}
\affiliation{Helmholtz-Zentrum Dresden–Rossendorf, Dresden, Germany}
\author{A. Jenkins} \author{L. C. Benetti}  \author{A. Schulman} \author{R. Ferreira}
\affiliation{International Iberian Nanotechnology Laboratory, Braga, Portugal}
\author{G. Philippe} \author{C. Chappert}
\affiliation{Universit\'e Paris-Saclay, CNRS, Centre de Nanosciences et de Nanotechnologies, Palaiseau, France}
\author{H. Schultheiss}  \author{K. Schultheiss} 
\affiliation{Helmholtz-Zentrum Dresden–Rossendorf, Dresden, Germany}
\author{J.-V. Kim} 
\affiliation{Universit\'e Paris-Saclay, CNRS, Centre de Nanosciences et de Nanotechnologies, Palaiseau, France}

\date{\today}                                           
                       
%%%%%%%%%%%%%%%%%%%%%%%%%%%%%%%%%%%%%%%%
%
%       Abstract
%
%%%%%%%%%%%%%%%%%%%%%%%%%%%%%%%%%%%%%%%%
\begin{abstract}
Forced excitations at frequencies in the range of the first order azimuthal spin waves of a magnetic disk in the vortex state are known to scatter into the vortex gyration mode, thereby allowing the growth of Floquet spin waves forming a frequency comb. We study the temporal emergence of this dynamical state using time-resolved microwave electrical measurements. The most intense Floquet mode emerges synchronously with the gyration mode after a common incubation delay which diverges at the scattering threshold. This delay is minimal when the drive is resonant with one of the first order azimuthal spin waves. It can be as short as 3 ns for the maximum investigated power. We conclude that the first-to-occur scattering mechanism is the three-wave splitting of a regular azimuthal eigenmode into a coherent pair formed by a gyration magnon and a Floquet spin wave.\\
\end{abstract}

\maketitle

%%%%%%%%%%%%%%%%%%%%%%%%%%%%%%%%%%%%%%%%
%
%                Paper
%
%%%%%%%%%%%%%%%%%%%%%%%%%%%%%%%%%%%%%%%%
%%%%%%%%%%%%%%%%%%%%%%%%%%%%%%%%%%

%\section{Introduction}
Magnetic disks having a vortex ground state have become a popular system both for magnonics applications \cite{yu_magnetic_2021} and as model test bench for micromagnetic theories \cite{shinjo_magnetic_2000, hertel_ultrafast_2007}. Their eigenspectrum comprises a low frequency mode --the translational motion of the vortex core, most often referred to as the gyration mode--, as well as the confined spin wave modes that have frequencies higher by typically more than one order of magnitude \cite{taurel_complete_2016}.  The high frequency modes are conveniently described by a radial index $n$ and an azimuthal index $m$, that respectively count the number of nodes along the disk radius and in the azimuthal direction along half a perimeter, respectively.  

The magnetization of vortex dots is easily set in the non-linear regime where spin waves start to scatter mutually \cite{schultheiss_excitation_2019, korber_nonlocal_2020, korber_pattern_2023, verba_theory_2021} and interact with the vortex core. The interactions between the core and spin waves have been studied along two main lines. The hybridization of the azimuthal spin waves with the gyration mode \cite{park_interactions_2005, guslienko_dynamic_2008} was first shown to lift the degeneracy between spin waves of opposite azimuthal indices \cite{park_interactions_2005, salama_large_2025}. In parallel, active research was conducted on the possibility to switch the polarity of the vortex core by interaction with azimuthal spin waves \cite{kammerer_magnetic_2011, yoo_azimuthal-spin-wave-mode-driven_2015}, sometimes in conjunction with an active pumping of the gyration \cite{sproll_low-amplitude_2014}. 

Together with these two regimes, interactions between the high frequency spin waves and the vortex core are occurring \cite{jenkins_electrical_2021}. This was studied numerically. Z. Gao et al. \cite{gao_interplay_2023} analyzed the reaction of the vortex core to an impinging spin wave beam. They showed that in addition to the forced excitation due to the incoming wave, core gyration is induced and the excitation spectrum splits 
into a frequency comb with finger spacing being the gyration frequency.  Z. Wang et al. \cite{wang_twisted_2022}  proposed that the physical origin of such frequency comb is the confluence and splitting scattering of $m=1$ azimuthal spin waves with the gyration mode of the vortex core.
The emergence of the frequency comb was then confirmed experimentally by C. Heins et al. \cite{heins_self-induced_2026, heins_control_2025}. The observed phenomenology --threshold behavior at which eigenmodes get suddenly highly populated-- bore a strong similarity with three-wave processes observed in other magnetic platforms \cite{boardman_three-_1988, bracher_parallel_2017}.
In addition, Heins et al. showed that the high frequency modes involved in the scattering are not the regular eigenmodes of the system but rather magnon Floquet states constructed on the time-periodic gyrating vortex, and populated by the nonlinear coupling of driven magnon modes to the core gyration.
These earlier works \cite{heins_self-induced_2026, heins_control_2025, philippe_excitation_2026} provided a clear understanding of the steady state situation and its event-averaged transient but did not dig into the dynamics of how azimuthal spin wave modes scatter into vortex core gyration and how they adapt to the resulting Floquet context. Since scattering is most often a stochastic process, some features of this process can only be revealed by real-time single-shot experiments. 

In this paper, we build on the works of ref.~\cite{heins_self-induced_2026, heins_control_2025, philippe_excitation_2026} and study the transient response of a vortex dot to in-plane rf magnetic fields of frequency close to that of the azimuthal spin wave modes. Time-resolved, high sensitivity measurements are enabled thanks to the embedding of the vortex dot in a magnetic tunnel junction.  
This allows the coincidence detection of the two magnon modes that split from the azimuthal spin waves, thereby demonstrating the three-wave nature of the scattering process. This elucidates the mechanism responsible for the generation of Floquet spin waves.

\begin{figure}%%%%%%%%%%%%%%%%%%%%%%%%%%%%%%%%%%
\includegraphics[width=8.4 cm]{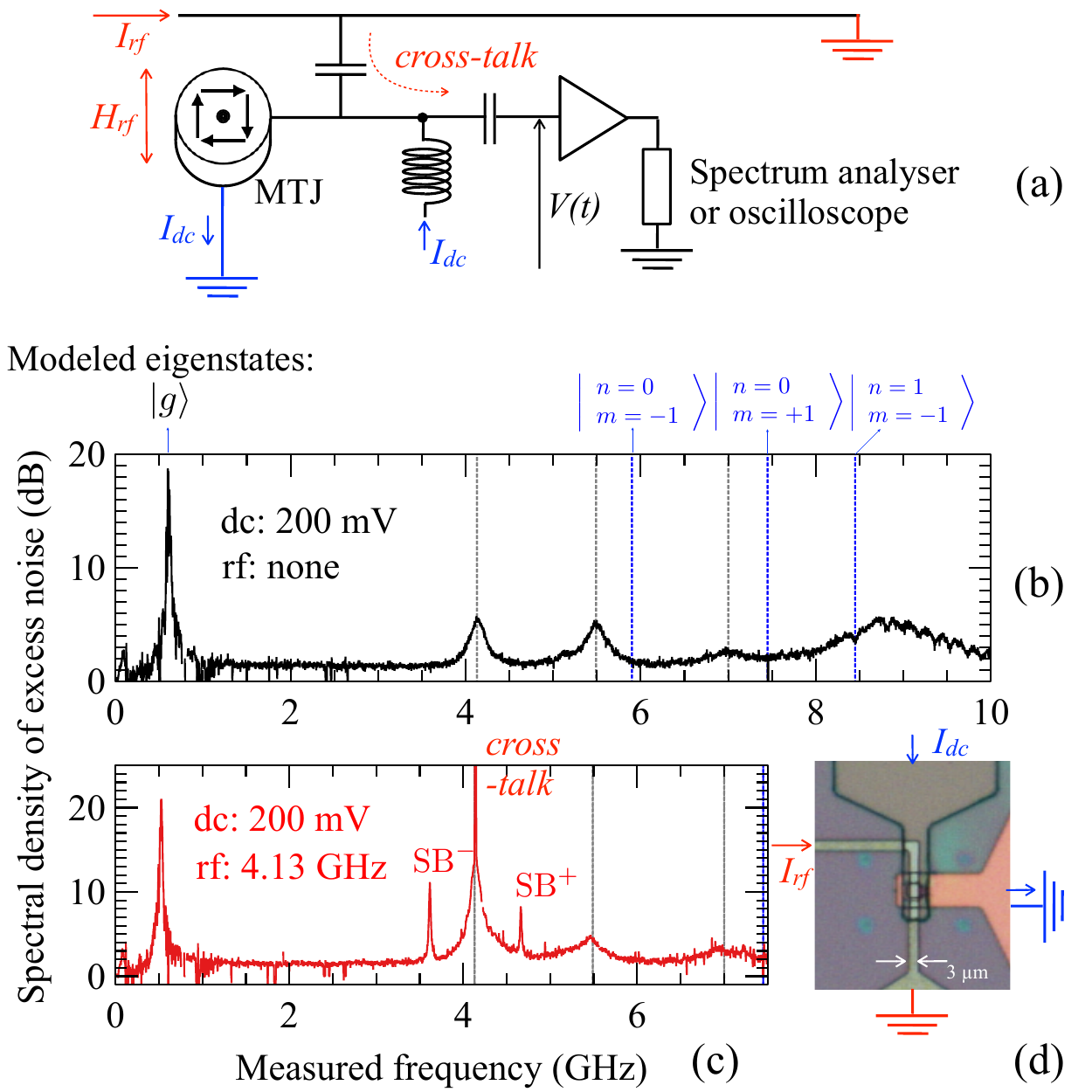}
\caption{Set-up and effect of the dc and rf drives in the linear regime. (a) A wire fed with an rf current applies a field on a circular magnet in the vortex state. The magnet is part of a magnetic tunnel junction which delivers a voltage $V(t)$ to the measuring instruments. (b) Thermal spin waves (grey lines): Increase of the power spectral density of $V$ when changing the dc bias from 0 to 200 mV. The blue lines are the frequencies of the modes of azimuthal indices $m=\pm 1$ in the micromagentic simulations. (c) Spectrum with the signal of thermal spin waves, cross-talk signal and microwave mixing signal leading to the sidebands SB$^-$ and SB$^+$ when applying an rf drive at 4.13 GHz and 10.8 dBm. (d): Optical micrograph of a device.}%_V1002_F1413_P1003, nominally 14 dBm
\label{Fig_Linear_Response}
% Data in NAS\Thib\Alex\03_2024_NL_SAandSCOPE\R11P04_MixedDomain_AWG\200mVbiasR11P04\Analyses\
% signal_theory_v4.opj
% Figure_Signal_theory_22_08_2025.key
%Figure1composite.key, Figure1Veuse.vsz

\end{figure}

%\section{Sample and setup}
We work with 45-nm-thick circular magnets made from a CoFeBSi alloy with a vortex ground state. We studied magnets of diameters ranging from 200 to 1000 nm, with typically 10 devices per diameter. The magnets are embedded in magnetic tunnel junctions (MTJs) of magneto-resistance 180\% and resistance of 320 $\Omega$ for a diameter of 300 nm. The reference layer is uniformly magnetized and fixed. The MTJs and their contact pads are small enough to be considered as lumped elements, with bandwidth exceeding 20 GHz.
 The spin fluctuations within the magnet are studied by sensing the voltage $V$ that the MTJ delivers to an amplifier equivalent to a $50~\Omega$ load [Fig.~\ref{Fig_Linear_Response}(a)]. This voltage is either analyzed in frequency domain using a spectrum analyzer  
 or time-resolved using a single-shot oscilloscope. We are sensitive to the spin excitations that result in variations $\delta R$ of the MTJ resistance. Because of the symmetry of our system, the largest $\delta R$'s correspond to magnetization fluctuations of first order azimuthal symmetry; this includes the vortex gyration and the spin waves with azimuthal indices \cite{ivanov_high_2005, boust_micromagnetic_2004} $m = \pm1$. Our sensitivity to other modes is small (but finite) and depends on the mode profile and how the space-resolved magneto-conduction dynamically redistributes the current across the MTJ as a response to this profile \cite{devolder_using_2017}.

Without any other excitation except the thermal fluctuations, $\delta R$ can already be revealed by dc biasing (current $I_\textrm{dc}$ from a bias of $\pm200$ mV) the MTJs and measuring the spectral density $||\tilde V(f)||^2$ of the voltage that the MTJ delivers [Fig.~\ref{Fig_Linear_Response}(b)]. This spectrum contains discrete lines well-described by the expected $\chi''(f) / f$ shape \cite{smith_modeling_2001} where $\chi''(f)$ is a Lorentzian function. The main fluctuation is at $597(26)~\textrm{MHz}$ and it is ascribed to the gyration  \cite{shreya_granular_2023}. The frequencies of the other fluctuators are $4.13(170)~\textrm{GHz}$, $5.49(240)~\textrm{GHz}$ and $7.0(\approx300)~\textrm{GHz}$, where the numbers in parentheses are the FWHM linewidths in MHz. We believe that the mode at 4.13 GHz belongs to the fixed layer of the MTJ and, as it will not react to the rf stimuli we shall disregard it. While the frequency of the gyration mode varies by $\pm 70~\textrm{MHz}$ from device to device as a result of material granularity \cite{jenkins_impact_2024}, the other ones are reproducible. Spin waves of higher frequencies are also thermally active; they lead to a broad spectral feature in the 8-10 GHz interval [Fig.~\ref{Fig_Linear_Response}(b)] that plays no role in our study. 

We performed \texttt{Magnum.np} eigenmode simulations \cite{bruckner_magnumnp_2023} using the measured magnetization ($M_s\approx500~\textrm{kA/m}$) and an estimated exchange stiffness of 10 pJ/m.  
The predicted eigenmodes include the gyration of frequency $f_\textrm{g}=585~\textrm{MHz}$, and confined spin wave modes of frequencies  $f_\textrm{n=0, m=-1}=5.9~\textrm{GHz}$, $f_\textrm{n=0, m=+1}=7.45~\textrm{GHz}$ and $f_\textrm{n=1, m=-1}=8.64~\textrm{GHz}$ mode. The sign of the azimuthal indices $m$ of the spin waves are given assuming a vortex core of positive polarity. The $\{n=0, m=+1\}$ mode has a thickness profile indicating some hybridization with the flexural gyration mode \cite{ding_higher_2014}. Small variations of the assumed material parameters and device size affect the frequencies of the modes, but not their frequency sequence. Owing to the reasonable numerical agreement, we assign the $\{n=0, m=-1\}$ and the $\{n=0, m=+1\}$ modes at 5.9 and 7.45 GHz to their experimental counterparts at 5.49 and 7.0 GHz. As from now on we shall only deal with $n=0$ modes, we will omit this index and quote them as $\lvert g \rangle$, $\lvert m=-1 \rangle$ and $\lvert m=+1 \rangle$.

%\section{Linear response}

Each MTJ is placed 1.2 $\mu\mathrm{m}$ below a 3 $\mu\mathrm{m}$-wide wire [Fig.~\ref{Fig_Linear_Response}(d)] connected to an rf source that supplies a fast rising ($<100~\textrm{ps}$) rf current $I_\mathrm{rf}$ of adjustable frequency $f_\textrm{rf}$ [Fig.~\ref{Fig_Linear_Response}(a)]. The generated rf field is of the order of 1 mT for 2.8 dBm of power arriving at the field line. It has three effects. \\ %6 dBm nominal 
(i) First, it generates a parasitic cross-talk responsible for the most intense peak in the $||\tilde V(f)||^2$ spectra, right at $f_\textrm{rf}$, as in the example of Fig.~\ref{Fig_Linear_Response}(c). 
\\ (ii) Second, $I_\mathrm{rf}$ generates an in-plane rf magnetic field. The layout of the sample ensures that the rf field is uniform at the scale of the device. 
This rf field is assumed not to actuate the magnetically-hard reference layers of the MTJ. Conversely, this rf drive induces a forced excitation of the magnet, excitation that we shall simply quote as $\lvert \mathrm{rf} \rangle$. By symmetry, the excitations with azimuthal symmetry $m = \pm1$ (including the gyration) are much more excited than any other ones \cite{boust_micromagnetic_2004}.
The created population of $\lvert \mathrm{rf} \rangle$ yields a resistance fluctuation $\delta R $ at the frequency $f_\textrm{rf}$. This should in principle be seen as an additional signal at $f_\textrm{rf}$ when a dc current biases the junction. Unfortunately the signal generated by this forced excitation is hidden below the cross-talk \footnote{except when $f_\textrm{rf}$ is directly resonant with the gyration mode, a situation that is not studied here}. So, the value $||\tilde V(f=f_\textrm{rf})||^2$ is unfortunately not magnetically informative. \\
(iii) Third, the interplay between the rf current and the magnetization fluctuations generates frequency mixing.
Indeed, if the resistance fluctuates by $\delta R_\textrm{fluct}$ at a frequency $f_\textrm{fluct}$ this modulates the amplitude of the MTJ voltage. This leads to two sidebands ($\mathrm{SB}^+$ and $\mathrm{SB}^-$) at the frequencies $f_\textrm{rf} \,\pm\, f_\textrm{fluct}$, each carrying a power: 
\begin{equation} P_\textrm{SB} \propto \delta R_\textrm{fluct}\,.\,I_\textrm{rf}^2 \label{mixing} \end{equation} 
and having a linewidth reflecting that of the fluctuator. In practice the gyration is \textit{always} (at least thermally) populated so that two mixing sidebands (SB$^-$ and SB$^+$) are bound to \textit{always} appear at $f_\textrm{rf} \pm f_\textrm{g}$. Fig.~\ref{Fig_Linear_Response}(c) is an illustration of this mixing-induced sidebands. For the parameters of Fig.~\ref{Fig_Linear_Response}(c), i.e. a 4.13 GHz drive, the rf field induces a linear response, so that the spectral signatures of the (thermally populated) regular eigenmodes just add to the effect of the drive [Fig.~\ref{Fig_Linear_Response}(c)]. These mixing-induced sidebands are still present if the dc bias is suppressed (not shown). 

We emphasize that in the linear regime this frequency mixing yields \textit{only the first} lateral sidebands while the cross-talk yields the peak at $f_\textrm{rf}$. The spectrum is thus trident-like [see Fig.~\ref{Fig_Linear_Response}(c)] and by no means this amplitude modulation effect could generate a frequency comb.

%\section{Steady state non-linear regime} %%%%%%%%%%%%%%%%%%%

We have stayed so far in a low amplitude regime for which there is no scattering among spin waves. Studies in comparable vortex disks \cite{heins_self-induced_2026} showed that when sufficiently populated, the spin waves scatter into the gyration mode, and in this case, the large gyration can create a Floquet context that renormalizes the spin wave manifold. Our setup can conveniently study the temporal evolution of this phenomenon. 

\begin{figure}%%%%%%%%%%%%%%%%%%%%%%%%%%%%%%%%%%
\includegraphics[width=8.4 cm]{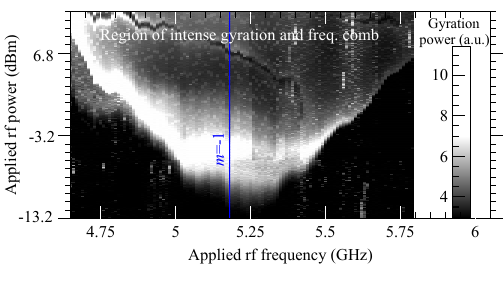}
\caption{(c) Total power of the gyration mode (defined as $\int_{10~\textrm{MHz}}^{1~\textrm{GHz}} ||\tilde V(f)||^2 df$) versus stimulus power and drive frequency $f_\textrm{rf}$ for a bias of 200 mV on a device of diameter 400 nm, whose $m=-1$ regular mode is at 5.18 GHz (vertical blue line). The black color level corresponds to the power generated by the thermal fluctuations of the gyration.}
\label{Fig_Tongue}
\end{figure}

\begin{figure}%%%%%%%%%%%%%%%%%%%%%%%%%%%%%%%%%%
\includegraphics[width=8.4 cm]{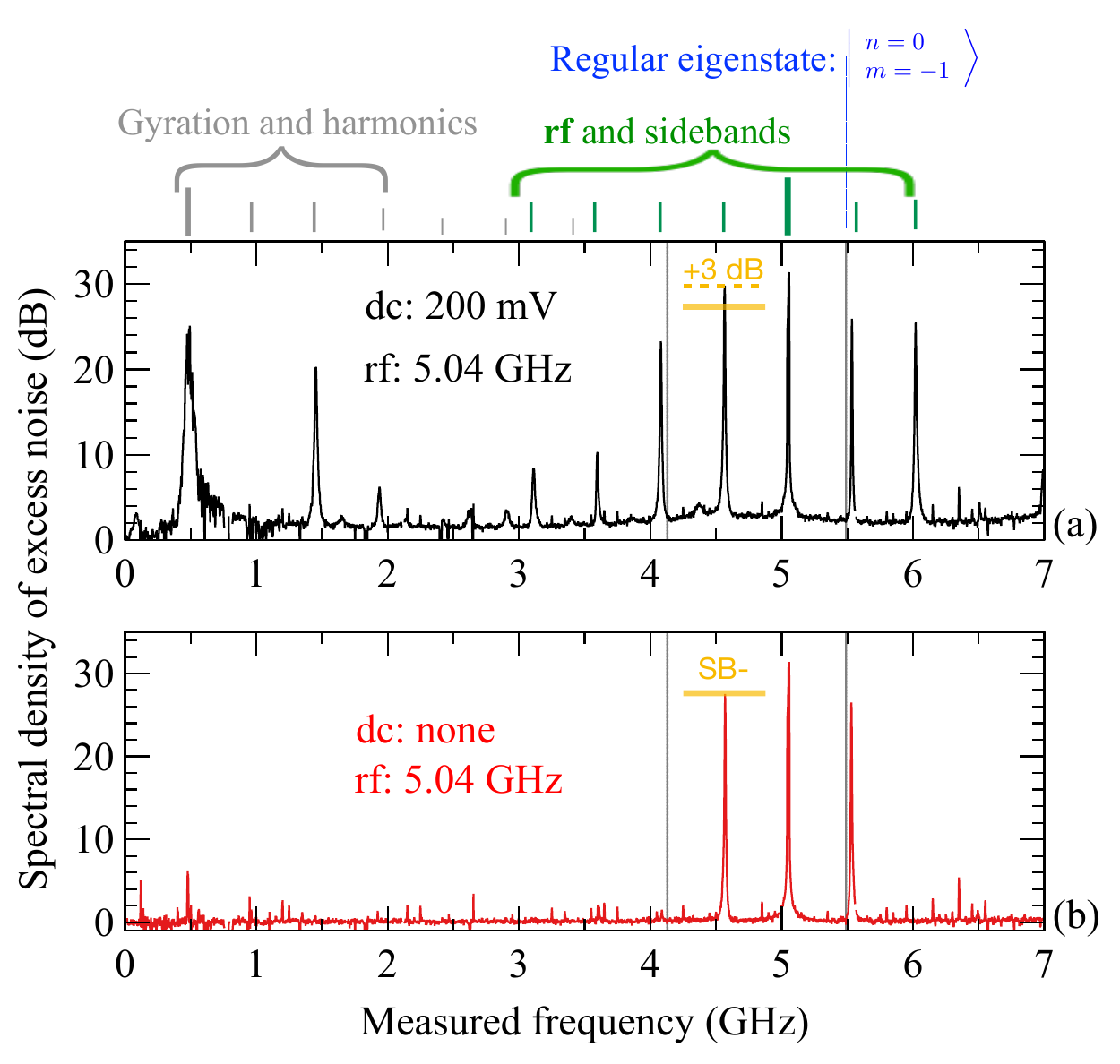}
\caption{Scattering of the forced $\lvert \textrm{rf} \rangle$ mode into the gyration mode and resulting frequency comb for a 300 nm diameter device. Power spectral density when applying a pump at 5.04 GHz with a power of 10.8 dBm (a) in the presence of 200 mV of dc bias, and (b) in the absence of dc bias. The dashed yellow bar is the power threshold to be used later when analyzing the time-resolved amplitude of the first lower sideband SB$^-$. } %nominally 14 dBm
\label{Fig_Frequency_comb}
\end{figure}

Fig.~\ref{Fig_Linear_Response}(c) showed that when $f_\mathrm{rf}$ was tuned to 4.13 GHz the forced mode $\lvert \textrm{rf} \rangle$ was not scattering. We confirmed that when the same conditions are used but $f_\mathrm{rf}$ is increased to a value closer to $f_{m=-1}$, the system abruptly transits into a different dynamical regime in which the device delivers a substantially larger power at the gyration frequency with profoundly modified spectral features at higher frequencies, including a frequency comb. A similar abrupt evolution happens if the rf power is increased. Meanwhile the coherence of the gyration also improves so that higher harmonics of the gyration frequencies appear. Fig.~\ref{Fig_Tongue} reports an example of the power emitted by the gyration (arbitrarily defined as $\int_{10~\textrm{MHz}}^{1~\textrm{GHz}} ||\tilde V(f)||^2 df$) versus the frequency $f_\textrm{rf}$ and power of the drive. The appearance of the strong gyration signal is accounted for by the well-defined bright tongue which is rather symmetric about $f_{m=-1}$. This confirms that the forced mode $\lvert \textrm{rf} \rangle$ scatters into the gyration mode ${\lvert g\rangle}$. Such transition was observed in all the investigated devices (i.e. for diameters in the 200 nm to 1 $\mu$m range) provided the drive frequency is tuned to their proper frequencies.  \\
Fig.~\ref{Fig_Frequency_comb}(a) shows that there appears concurrently an additional frequency comb centered about the drive frequency and spanning the values $f_\textrm{rf} + k f_\textrm{g}$, with $k \in \mathbb Z$.  The sideband $\textrm{SB}^-$ (i.e. $k=-1$) is systematically the most intense one after the peak at the drive frequency.  This could have been expected since the scattering of the forced mode into the gyration mode must generate a companion quasi-particle of energy $f_\textrm{rf}-f_\textrm{g}$ to conserve energy. We conclude that the first-to-occur scattering mechanism is thus very likely to be:
\begin{equation} \underbrace{\lvert \textrm{rf}\rangle}_{m=-1}  \xrightarrow[\textrm{splitting}]{\textrm{ }} \underbrace{\lvert g\rangle}_{m=+1} + \, \underbrace{\lvert \textrm{SB}^- \rangle}_{m=-2}~,
\label{ScatteringReaction} %%%%%
\end{equation}
where the azimuthal indices of the states $\lvert \textrm{rf} \rangle$ and $\lvert \textrm{SB}^- \rangle$ will be deduced later from  conservation laws \cite{verba_theory_2021}. Note that the splitting is very far from degenerate, which is an unusual situation in non-linear magnonics \cite{kim_stimulated_2024}.

In related studies \cite{heins_self-induced_2026, heins_control_2025, philippe_excitation_2026} the frequency comb has been interpreted as the signature of Floquet states residing on a time-periodic gyrating vortex, with the frequencies reflecting the quasi-energies of the Floquet states. 
This interpretation is supported by two of our findings:\\(i) At the crossing of the scattering threshold, the power of the gyration mode increases substantially (Fig.~\ref{Fig_Tongue}). The ground state on which the magnetic excitations reside ceases to be a non-coherently (thermally-driven) gyrating vortex and becomes a time-periodic gyrating vortex, supplying a Floquet context to the other eigenexcitations. \\(ii) Correlatively, the regular eigenmodes of the static vortex that were easy to detect in sub-threshold regime (Fig.~\ref{Fig_Linear_Response}) disappear suddenly [Fig.~\ref{Fig_Frequency_comb}(a)] and are replaced by the new spectral features with different frequencies, now being $f_\textrm{rf} + k f_\textrm{g}$, $k \in \mathbb Z$. 

It is essential to assess whether the novel spectral features at $f_\textrm{rf} + k f_\textrm{g}$ --the directly measured peaks-- correspond to \textit{real} states --the inferred modes--, or if these spectral features simply result from microwave mixing phenomena. This question is conveniently answered by setting $I_\textrm{dc}=0$ because this suppresses the direct sensitivity to the modes while maintaining the mixing phenomena. Fig.~\ref{Fig_Frequency_comb}(b) confirms that the gyration peak and the sidebands for $|k| > 1$ entirely \footnote{In fact, the rectification of $I_\textrm{rf}$ by a synchronous variation of the device resistance generate a small dc current flowing though the MTJ, which renders the gyration observable in the spectra as a small narrow line at 480 MHz.} disappear from the spectra when the dc bias is suppressed. This proves that the spectral lines for $|k|>1$ correspond to true states and that these states are populated. Conversely, the first lateral sidebands (i.e. $k=\pm1$) do not vanish when the dc bias is suppressed: $\textrm{SB}^-$ just decreases by 3 dB (see the yellow bars in the two panels of Fig.~\ref{Fig_Frequency_comb}). It decreases but does not vanish simply because the large gyration maintains a large mixing signal at this frequency (Eq.~\ref{mixing}). Still, this reduction of the $|k| = 1$ sideband amplitude is a proof of the existence of populated \textit{real} states also for $k=\pm1$.

\begin{figure}%%%%%%%%%%%%%%%%%%%%%%%%%%%%%%%%%%
\includegraphics[width=8.4 cm]{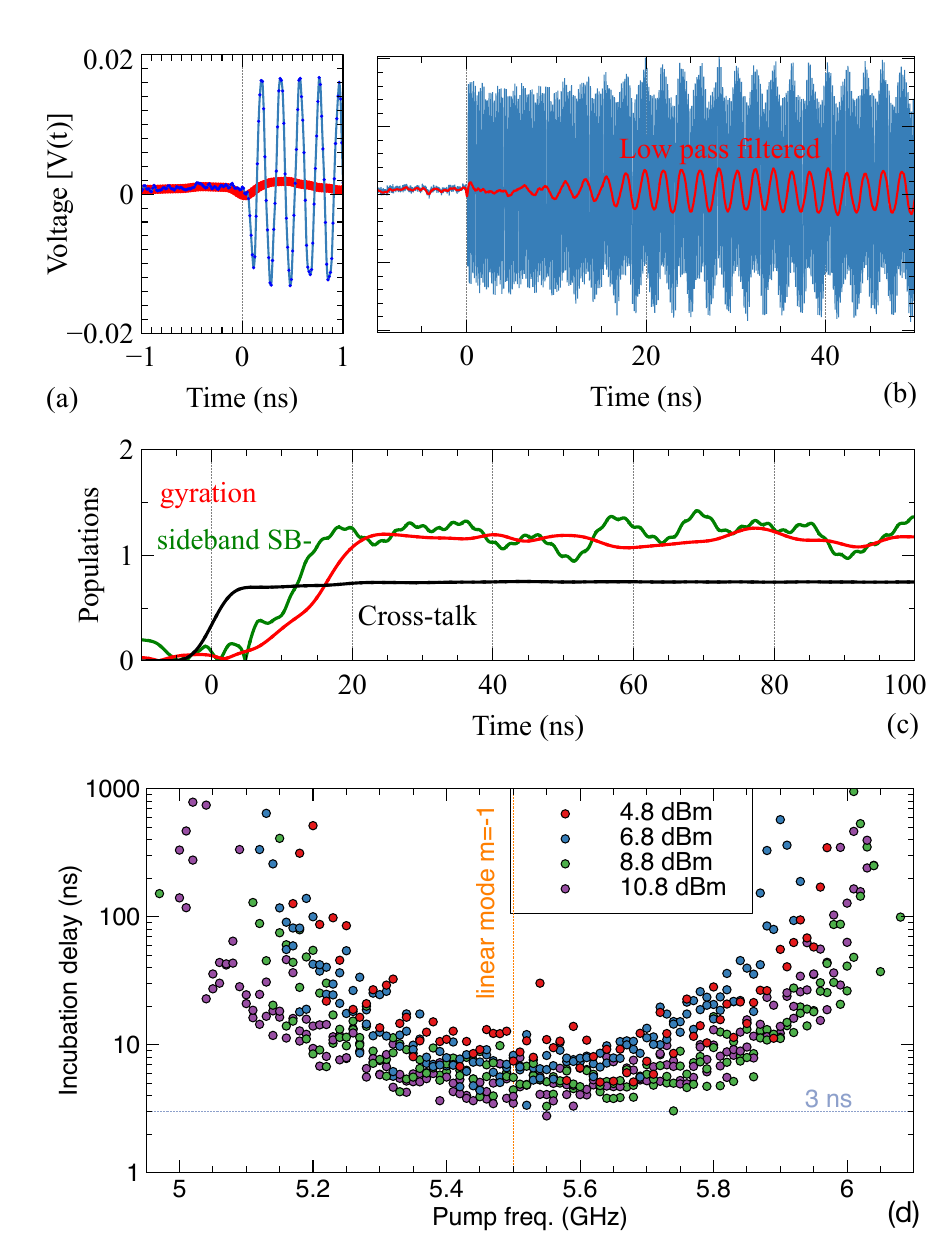}
\caption{Time-resolved population of the gyration mode for a 300 nm device. (a, b) $V(t)$ (blue) and a low-pass-filtered version of it (red) for a drive at $f_\textrm{rf}=5.17$ GHz and 8.8 dBm with a dc bias of -200 mV.  %nominally 12 dBm
(c) Transients of the populations $n(t)$ of the gyration mode at $f_\textrm{g}=492$ MHz (red, $\times 8$ magnification), of the Floquet mode of the lower sideband SB$^-$ at $f_\textrm{rf}- f_\textrm{g}$ (green, $\times 60$ magnification) and of the signal at the forcing frequency $f_\textrm{rf}$ (black). The oscillations in the population of SB$^-$ originate from the experimental noise. (d) Dependence of the incubation delay of the gyration versus the pump frequency and the pump power. The vertical line is at the frequency of the regular eigenmode $\lvert m=-1 \rangle$.}
\label{FigDemodulation}
% V1002_F1516_P1003.tsv
% Data in NAS\Thib\Alex\03_2024_NL_SAandSCOPE\R11P04_MixedDomain_AWG\200mVbiasR11P04\Analyses\
% Illustration_Demodulation_v2.vsz
\end{figure}

% \section{Time-resolving the scattering process}
We now discuss the dynamics of the scattering of the spin waves into the gyration (Eq.~\ref{ScatteringReaction}). For this we replace the spectrum analyzer by a single-shot oscilloscope [Fig.~\ref{Fig_Linear_Response}(a)] and analyze the transients of $V(t)$ after an abrupt application of the rf drive in the presence of a dc bias. A representative example of $V(t)$ is supplied in Fig.~\ref{FigDemodulation}: in the first ns, $V(t)$ almost only reflects the cross-talk due to the fast-rising rf field at $f_\mathrm{rf}$. Later, a smaller signal at the gyration frequency can be perceived as an added slower oscillation. To follow the population of the gyration and of the other modes, we performed a numeric I/Q demodulation of $V(t)$. The population $n_f$ of a mode at a given frequency $f$ scales with the modulus of the down-converted signal:
\begin{equation}n_f(t) \propto \big\lvert \big\lvert \mathcal{F}_\textrm{low} \left(V(t) e^{i 2 \pi f t}\right)  \big\lvert \big\lvert \label{IQmethod}
\end{equation} 
where $\mathcal{F}_\textrm{low}$ is a delay-free low-pass filter of cut-off frequency $f_\textrm{g}$. This value enables the separate tracking of the populations responsible for the individual sidebands, while maintaining a time resolution of $ \pi / ( 2f_\textrm{g}) \approx3~\textrm{ns}$ (see appendix A).
This demodulation procedure degrades the time resolution of the populations $n_f(t)$ compared to that of the raw data $V(t)$, as evident when looking at the rise time of the demodulated amplitude of the cross-talk-dominated signal at $f_\mathrm{rf}$ [Fig.~\ref{FigDemodulation}(c), black curve](see appendix C for more details).

\begin{figure}%%%%%%%%%%%%%%%%%%%%%%%%%%%%%%%%%%
\includegraphics[width=8.4 cm]{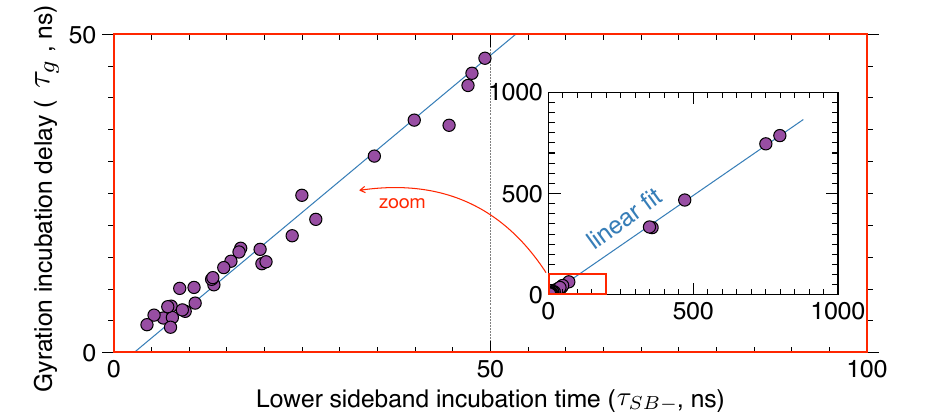}
\caption{Correlation between the incubation delays of the gyration and of the appearance of the signal of the lower sideband. Each point is a single shot event for applied frequencies spanning from 5 to 5.44 GHz, a power of 10.8 dBm and dc biases of -200 and +200 mV. } %nominally 14 dBm
\label{Fig_IncubationDelays}
% Files V1002_F15*_P1003.tsv and .pops
% Correlation_Incubation_Times.vsz
% Data in /Volumes/nomadenas/Thib/Alex/03_2024_NL_SAandSCOPE/R11P04_MixedDomain_AWG/200mVbiasR11P04/Analyses/Delays_compilation/Correlation_Delays_Gyro_sideband.opj
\end{figure}

We clearly observe that a delay is needed for the gyration to reach its steady state amplitude [see Fig.~\ref{FigDemodulation}(a)]. We thus define the "incubation delay" $\tau_\textrm{g}$ of the gyration as the time required to reach within 3 dB of its average steady state value. This delay $\tau_\textrm{g}$ fluctuates stochastically and can occasionally be as short as 3 ns [Fig.~\ref{FigDemodulation}(d)].  Although in oscilloscope measurements the limited acquisition time window prevents extracting the exact form of the divergence, we can state from spectrum analyser measurements that $\tau_\textrm{g}$ diverges (i.e. exceeds 10 s) when reaching the border of the scattering region of the $\{P_\textrm{rf}, f_\textrm{rf}\}$ space, as earlier displayed in Fig.~\ref{Fig_Tongue}. 

The mean value of $\tau_g$ is clearly minimal when $f_\textrm{rf} = 5.5~\textrm{GHz}$, i.e. when the drive is resonant with the regular eigenmode $m=-1$ [see Fig.~\ref{FigDemodulation}(d)]. The minimal $\tau_g$ is 3 ns; this value is a true delay, not an artefact of the demodulation procedure (see appendix A). This dependence of $\tau_g$ with the carrier frequency bears some similarity with $1/\chi''(f)$, i.e. the inverse of the susceptibility of the $m=-1$ mode. This minimum of $\tau_g$ at $f_{\lvert m=-1 \rangle}$ indicates that the quasi-particle $\lvert \textrm{rf} \rangle$ that scatters into the gyration mode is an off-resonantly excited $\lvert m=-1 \rangle$ regular eigenmode, confirming the speculation of Eq.~\ref{ScatteringReaction} and the micromagnetic simulations of ref. \cite{philippe_excitation_2026}. 
The forced mode $\lvert \textrm{rf} \rangle$ that scatters and the gyrotropic mode $\lvert g\rangle$ thus share the opposite azimuthal numbers.
Since the total azimuthal number must be conserved in the scattering process \cite{verba_theory_2021}, this scattering picture entails that the companion mode $\lvert \textrm{SB}^- \rangle$ is a Floquet state of azimuthal character $m=-2$.

In the scattering picture of Eq.~\ref{ScatteringReaction}, the companion quasi-particle of energy $f_\textrm{rf}-f_\textrm{g}$ should be emitted simultaneously with the gyration magnon. Let us check this by time-resolving the population of the Floquet mode $k=-1$ through the I/Q demodulation of the sideband $\textrm{SB}^-$. 
The $n_\textrm{SB-}(t)$ curves [Fig.~\ref{FigDemodulation}(c)] resemble the ones of the gyration: they are step functions with incubation delays (see appendix B for a statistical view of the time-resolved populations). To ensure that their incubation delays $\tau_\textrm{SB-}$ reflect the \textit{cumulated} growth of both the mixing part of the signal and the contribution of the population of the $k=-1$ Floquet mode, $\tau_\textrm{SB-}$ is defined with an amplitude threshold set at 2 dB below our best estimation of the average steady state level of $n_\textrm{SB-}$, chosen as $\langle n_{SB-}(t \gg \tau_\textrm{SB-}) \rangle$. This threshold was earlier displayed as the yellow dashed bar in Fig.~\ref{Fig_Frequency_comb}: it is above the contribution of the sole microwave mixing part. Other choices for the threshold would lead to similar conclusions (see appendix A).

Despite the low signal to noise ratio and the resulting degraded time resolution, we can state that the growths of $n_\textrm{g}$ and $n_\textrm{SB-}$ are synchronous and share a \textit{common} incubation delay. To better evidence that despite its event-to-event variability this is a common delay, we show in Fig.~\ref{Fig_IncubationDelays} that there is a one-to-one correlation between the independently extracted $\tau_\textrm{g}$ and $\tau_\textrm{SB-}$.  
The $k=-1$ Floquet mode and the gyration mode are born simultaneously, such that the scattering is in fact a thee-wave splitting. 

In summary, our experiments demonstrate that when pumping a vortex dot with an in-plane rf magnetic field, the first-to-occur scattering mechanism is a splitting of a $\lvert m=-1 \rangle$ regular eigenmode into the pair formed by a gyration magnon and a Floquet spin wave. The single-shot time-resolved observation of the coincidence of appearance of the two products of the scattering demonstrate that it is a three-wave process.
The scattering threshold and the scattering time are minimal when the $\lvert m=-1 \rangle$ regular eigenmode is resonantly excited.
The appearance of the Floquet context related to the gyrating ground state and of the Floquet spin wave is thus not an egg-and-chicken problem: the two are generated synchronously by the first-to-occur scattering process. The subsequent time evolution of the system is an unresolved problem that will deserve attention in the future as the first splitting likely initiates a cascade of non-linear interactions. The primary products probably seed further scattering processes that progressively populate the other sidebands. 
Besides, it would interesting to determine the path taken by the firstly driven regular eigenstate to later evolve into, or be replaced by a Floquet mode of same azimuthal number when in the steady state regime.

\begin{acknowledgments}
We acknowledge financial support from the EU Research and Innovation Programme Horizon Europe under grant agreement n°101070290 (NIMFEIA). We thank L. Körber and J. H. Mentink from Radboud University for insightful discussions.
\end{acknowledgments}

\section*{DATA AVAILABILITY}
The data that support the findings of this article are openly available \cite{devolder_datasets_2026}.

\section*{Appendices}
\subsection*{Appendix A: Reliability of the time metrics}

\begin{figure}%%%%%%%%%%%%%%%%%%%%%%%%%%%%%%%%%%
\includegraphics[width=8 cm]{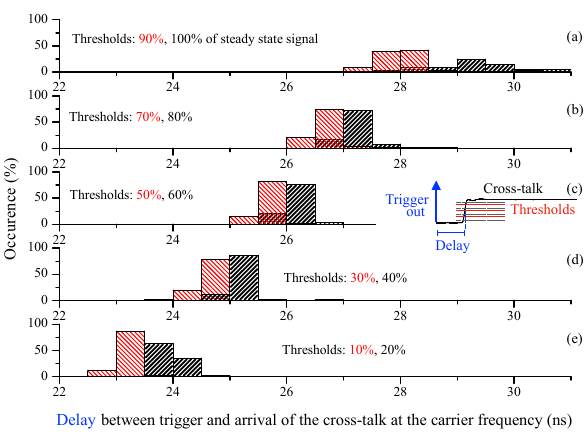}
\caption{Characterization of the precision of the time of onset of the stimulus. Inset: sketch of the definition method. The oscilloscope starts the measurement when receiving the trigger out signal sent by the arbitrary waveform generator. The MTJ signal, which is dominated by the cross-talk, arrives circa 25 ns later because of the trigger delay and of group delay of the cables and amplifiers in the measurement path. We define the instrument delay times by setting a software threshold of a variable fraction (10\% to 100\%) of the steady state signal $\langle n_\textrm{rf} \rangle$. Panels (a-e): histograms of the instrument delay times for variable choices of the threshold levels.} 
%nominally 14 dBm
%/Volumes/nomadenas/Thib/Alex/03_2024_NL_SAandSCOPE/R11P04_MixedDomain_AWG/200mVbiasR11P04/TSVs/P1002/PopVersusTime/PopVersusTime_F1500toF1599/IncubDec2025_GYRO_V1000/InfluenceOfDecileOnErrorIn_Xtalk_Incub.opj
\label{Fig_InstumentDelay}
\end{figure}

\begin{figure*}%%%%%%%%%%%%%%%%%%%%%%%%%%%%%%%%%%
\includegraphics[width=14 cm]{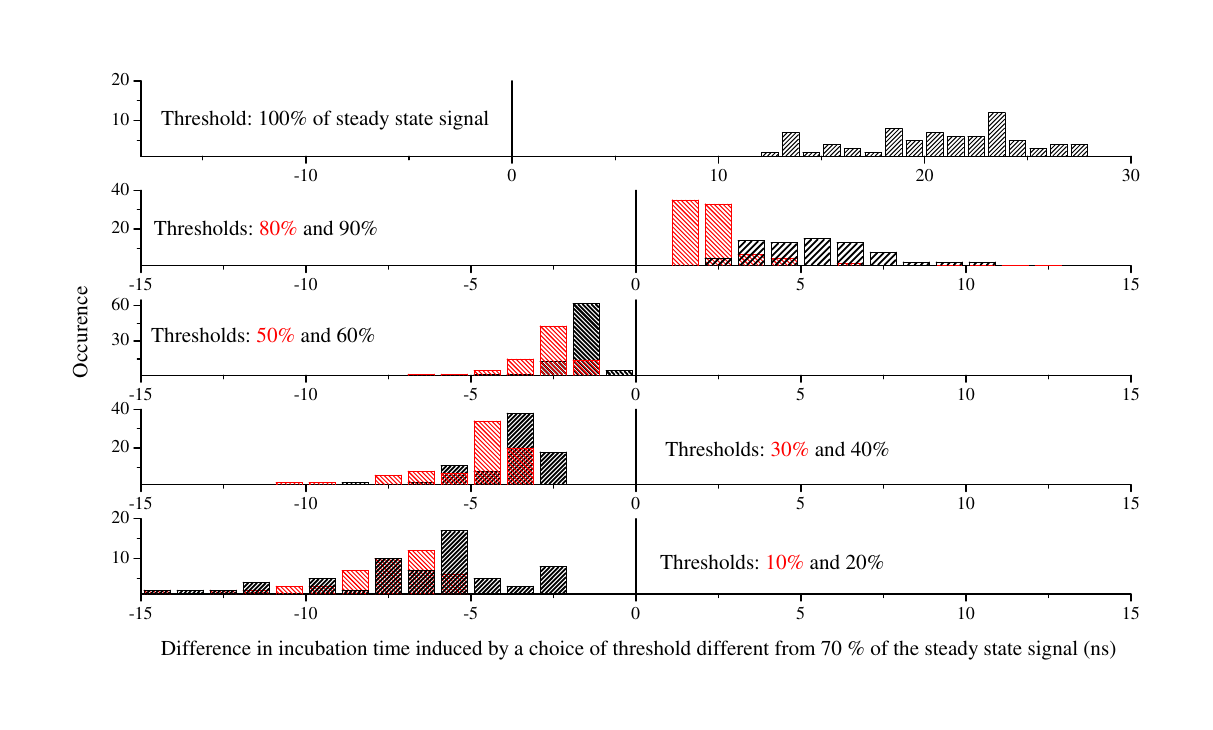}
\caption{Characterization of the error that would be done on the incubation time $\tau_g$ of the gyration mode if other values of  threshold were chosen. The reference is a software threshold of taken at 70\% of the steady state signal $\langle n_\textrm{g} \rangle$ measured after the splitting. The top two panels gather the error histograms for thresholds larger than the 70\% reference, and the bottom panels are for lower thresholds.} 
 %nominally 14 dBm
%/Volumes/nomadenas/Thib/Alex/03_2024_NL_SAandSCOPE/R11P04_MixedDomain_AWG/200mVbiasR11P04/TSVs/P1002/PopVersusTime/PopVersusTime_F1500toF1599/IncubDec2025_GYRO_V1000/InfluenceOfDecileOnErrorInGyroIncub_v3.opj
\label{Fig_IncubationTime}
\end{figure*}

Fig.~\ref{Fig_InstumentDelay} and Fig.~\ref{Fig_IncubationTime} illustrate the precision of the definition of the time metrics. They illustrate the robustness of the determination of the incubation time depending on the chosen threshold, and illustrates the influence of the low pass filter on the definition of the incubation times.

%the 5 ns of the low pass filter.
The demodulation procedure (Eq. 3) is applied on 100 randomly chosen time-traces to deduce (i) the time at which the stimulus is applied with respect to the trigger (Fig.~\ref{Fig_InstumentDelay}) and (ii) the incubation times of the gyration signal (Fig.~\ref{Fig_IncubationTime}) and of the SB- signal (Fig. S3). In each case determine the time at which the demodulated amplitude crosses variable thresholds. The investigated thresholds are chosen as fractions (10\%, 20\%,...,100\%) of the steady state signal level $\langle n_\textrm{rf} (t)\rangle$, which we define as the time-averaged demodulated amplitude on a time window from 950 ns to 1 $\mu$s after starting the pump. 

The steady-sate level is estimated from the sole curves where the splitting is observed. The recorded curves that do not include a scattering event in that time interval are easily discarded because they lead to step-less time-resolved curves similar to the one in Fig. S4(f). They can be automatically identified by their low signal which obeys the significance-above-noise criterion $\frac{n_\textrm{rf}(t)}{\delta n_\textrm{rf}} < 5 $, $\forall t \in [0.95, 1 ~\mu \textrm{s}]$. $\delta n_\textrm{rf}$ is a measurement of the noise flow and we define it as the standard deviation of $n_\textrm{rf}$ measured for $t<0$.

For the cross-talk, the SNR is very large and the time at which the stimulus arrives at the sample is always found with a precision greater than 0.5 ns, except if an unreasonable threshold of 100\% is chosen (see the width of the histograms in Fig.~\ref{Fig_InstumentDelay}). The MTJ signal arrives circa 25 ns after the trigger signal sent by the arbitrary waveform generator flagging the onset of the microwave stimulus at $\omega_\textrm{rf}$. This delay is the combination o the trigger delay and the group delay of the cables and amplifiers in the measurement path. This instrumental delay is corrected for in all reported measurements, except in Fig.~\ref{Fig_InstumentDelay}.  
The crossing of the threshold is delayed by 5 ns if the threshold is increased from 10\% to 90\%. This reflects the slew rate of the low pass filter used in the demodulation procedure (Eq. 3). 

All the reported results in the main manuscript were done by setting a threshold at -3 dB (70\%) of the steady state signal of the cross-talk and of the gyration, and -2 dB for the lower sideband. It is important to inquire how this choice of threshold affects the definition of the incubation delays. Fig.~\ref{Fig_IncubationTime} studies this point for the definition of the gyration incubation delays $\tau_g$ and its statistics. This is done on a set of randomly chosen time traces $n_\textrm{g}(t)$ from which we have removed the ones that do not exhibit three-magnon splitting (the non-switching time traces are easily identified since in this case $\frac{\langle n_\textrm{g} \rangle}{\delta n_\textrm{g}} \leq 5$).
Fig.~\ref{Fig_IncubationTime} reports the histograms of the error in $\tau_g$ done by changing the threshold away from 70\% of the steady state signal. By definition, this error vanishes for 70\% and is therefore not reported in Fig.~\ref{Fig_IncubationTime}. Since the SNR of the $n_\textrm{g}(t)$ curves is much smaller than that of the $n_\textrm{rf}(t)$ curves, the histograms of the $\tau_g$'s have a width that is much larger than the formerly seen 0.5 ns time precision for the cross-talk signal. The width of the histograms is a couple of ns for threshold in the 30\% and 80\% interval. The modest SNR results in a substantial degradation of the precision of $\tau_g$ if the threshold is taken either below 20\% [Fig.~\ref{Fig_IncubationTime}(e)] or above 90\% [Fig.~\ref{Fig_IncubationTime}(a)]. In the range of the 30\% and 80\% interval of threshold, the mean value of the error in $\tau_g$passes from -4 ns to +2 ns, which can be understood from the slew rate of the low pass filter used in Eq. 3; if taken within that a 20-80\% range, the exact choice of the threshold does not influence our overall conclusions. A slightly larger threshold of 80\% (i.e. -2 dB) is chosen for the determination of $\tau_{SB-}$ for a better immunity to the artefactual spectral leakage of $n_\textrm{rf}(t)$ at the lowest end of the distribution of incubation times (see section III).

\begin{figure*}%%%%%%%%%%%%%%%%%%%%%%%%%%%%%%%%%%
\includegraphics[width=12. cm]{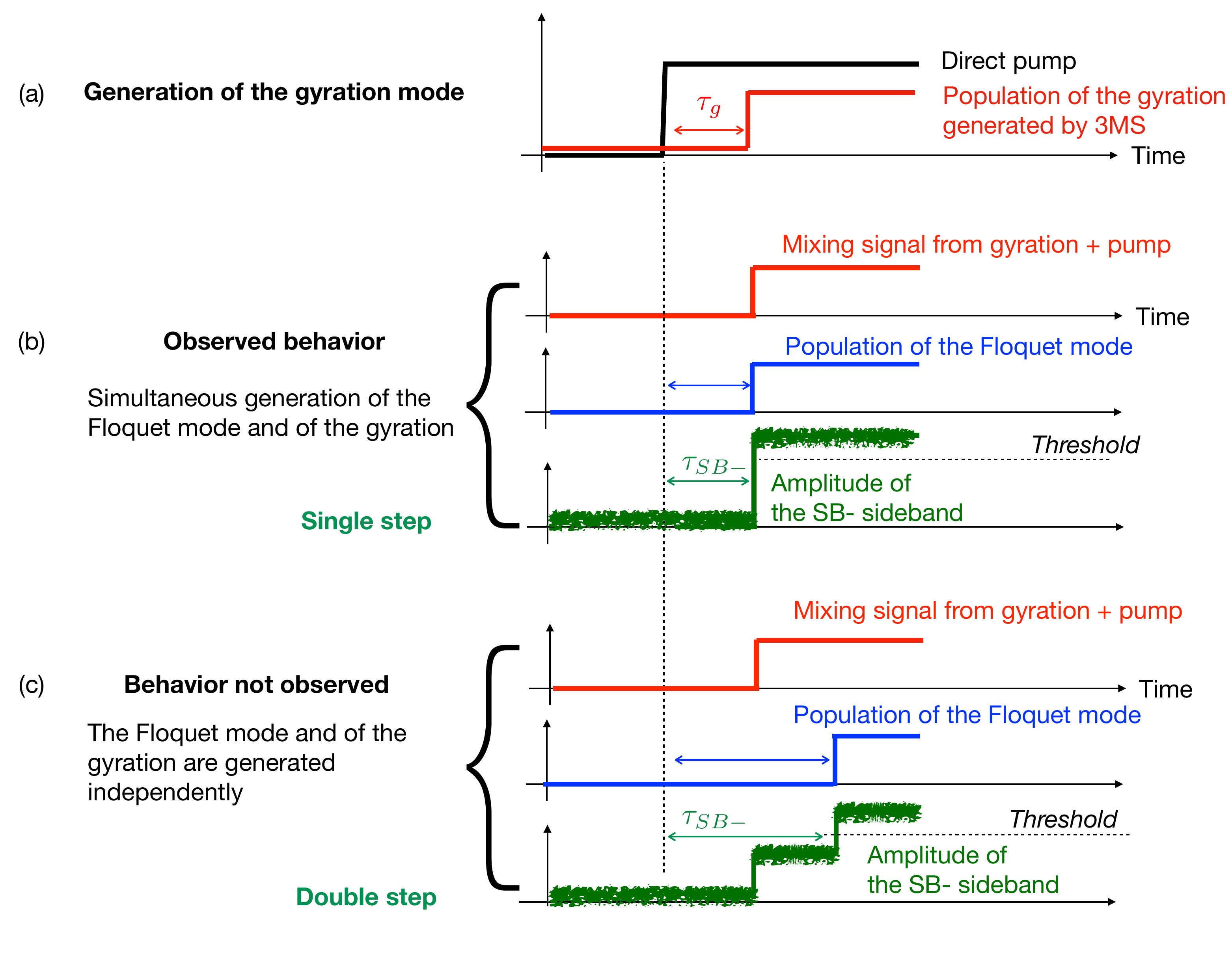}
\caption{(a): Qualitative shapes expected for the cross-talk signal $n_\textrm{rf}(t)$ resulting from the direct pump (black) and for the population of the gyration mode $n_\textrm{g}(t)$ (red).  (b, c): shapes expected for the signal due to the mixing of the rf stimulus and the gyration (red), for the population of the Floquet mode (blue) and the resulting signal of the lower sideband SB$^-$ (green). Panel (c) assumes that the Floquet mode is generated by three-magnon splitting splitting according to Eq. 2, whereas panel (c) assumes that the Floquet mode is generated by another process. } %nominally 14 dBm
%/Users/thibaut.devolder/Nextcloud/ARTICLES_soumis/2025_SWsgenerateGyro/Data_for_Figures/Supplementary/SingleSteppedOrNOT_v2.key
\label{Fig_SingleStepOrDoubleStep}
\end{figure*}

\subsection*{Appendix B: Separating the contributions of microwave mixing and Floquet band population to sideband amplitude}

\begin{figure*}%%%%%%%%%%%%%%%%%%%%%%%%%%%%%%%%%%
\includegraphics[width=12.4 cm]{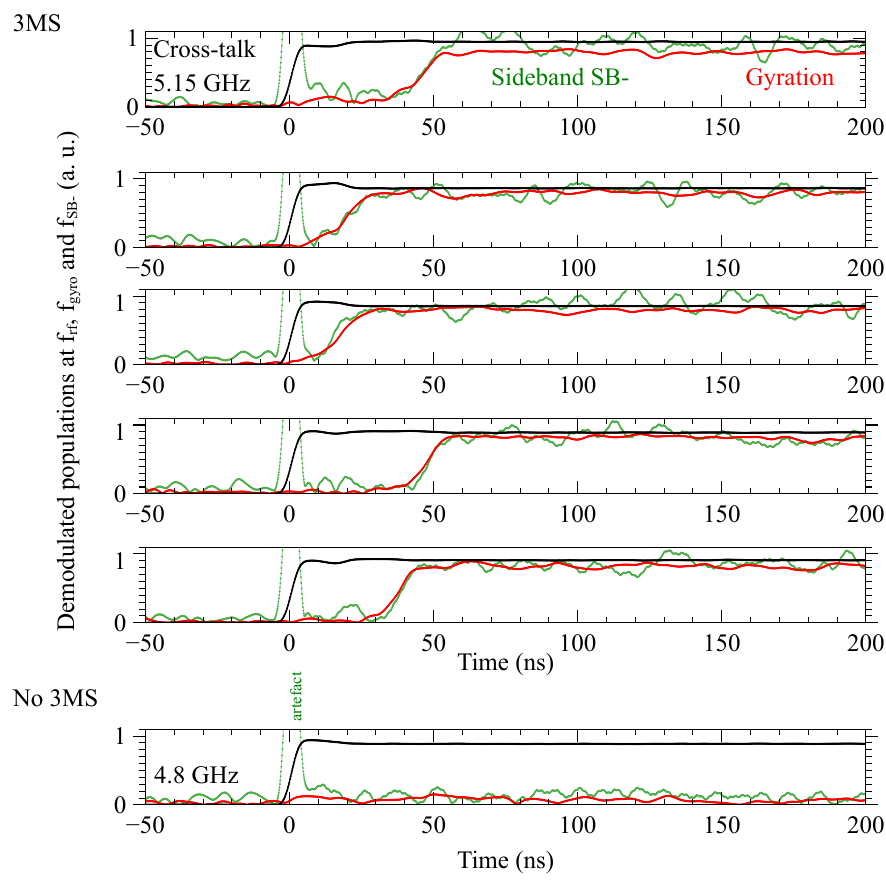}
\caption{Single-shot time-resolved transients of the populations $n(t)$ of the gyration mode (red), of the Floquet mode of the lower sideband SB$^-$ (green) and of the signal at the forcing frequency $f_\textrm{rf}$ (black). Five top panels: for a forcing frequency of 5.15 GHz and a power of 10.8 dBm that leads to three-magnon splitting (3MS). Bottom panel: for a forcing frequency of 4.8 GHz and a power of 10.8 dBm, that does not lead to three-magnon splitting. The overloading peak in $n_\textrm{SB-}(t)$ at $t\approx 0$ is due to the spectral leakage from the much more intense $n_\textrm{rf}(t)$ signal.
% /Users/thibaut.devolder/Nextcloud/ARTICLES_soumis/2025_SWsgenerateGyro/Data_for_Figures/Supplementary/TimreResolvedManyCurves_Feb2026.vsz
 } %nominally 14 dBm
\label{Fig_TimeResolvedTransients}
\end{figure*}

As mentioned in the main text, it is essential to assess whether the novel spectral feature at $f_\textrm{rf} - k f_\textrm{g}$ corresponds to a \textit{real} state (the Floquet mode) with the same frequency, or if this spectral feature simply results from microwave mixing between the applied rf and the gyration. This question was answered in the main manuscript by setting $I_\textrm{dc}=0$ and suppressing the direct sensitivity to the modes while maintaining the mixing phenomena. Here we revisit this point in Figs.~\ref{Fig_SingleStepOrDoubleStep} and \ref{Fig_TimeResolvedTransients}

Fig.~\ref{Fig_SingleStepOrDoubleStep} shows the expected qualitative shape of the time-resolved populations in two hypothetical scenarios. The middle panel shows the expected shapes of $n_\textrm{rf}(t)$ (black), $n_\textrm{g}(t)$ and $n_\textrm{SB-}(t)$ (green) if the Floquet and the gyration mode appear synchronously as hypothesized in Eq. 2 and concluded from the main manuscript. The bottom panel shows the shapes expected if the two modes were appearing at different instants, i.e. in disagreement with Eq. 2.
If Eq. 2 was correct, the mixing contribution (red) to the amplitude of the SB$^-$ sideband and the Floquet mode contribution (blue) to this same sideband are synchronous, such that $n_\textrm{SB-}(t)$ (green) is expected to be a single-stepped curve, as indeed observed experimentally. Conversely if these two contributions were not synchronous, the  $n_\textrm{SB-}(t)$ curve would be a two-stepped function with only one of the two steps  coinciding with the rise of $n_\textrm{g}(t)$. From Fig.~3 (see the yellow bars), the height of the second step would typically amount to a third (i.e. -3 dB) to the total signal.

The Fig.~\ref{Fig_TimeResolvedTransients} reports five randomly chosen (but representative) sets of time-resolved signals $n_\textrm{rf}(t)$ (black), $n_\textrm{g}(t)$ (red) and $n_\textrm{SB-}(t)$ (green). They show that the $n_\textrm{SB-}(t)$ is always single-stepped, and that the step coincides with that of the gyration signal $n_\textrm{g}(t)$. This is an additional proof that the spectral line SB$^-$ correspond to the population of a true state that is populated synchronously with the gyration.

\subsection*{Appendix C: Signal processing steps}
The voltage $V(t)$ delivered by the MTJ is amplified by a low noise amplifier of 3 dB analog bandwidth being 10 MHz to 15 GHz. Its noise figure is 5 dB at frequencies above 100 MHz. For the frequency-domain measurements, the amplified voltage is sent to spectrum analyser of resolution bandwidth 2 MHz. 

For the time-domain measurements, a single-pole-double-through router is activated to send the amplified voltage to an oscilloscope of sampling rate of 80 GSamples/sec. The analog bandwidth of the instrument is 25 GHz. Another channel of the oscilloscope is used to record the stimulus and to mathematically construct a noise-free complex harmonic signal at $f_\textrm{rf}$ that will be used for the demodulation of the $n_\textrm{rf}(t)$ using Eq. 3.  For each applied $f_\textrm{rf}$, we reconfirm the value of $f_\textrm{g}$ from the spectral analysis of the $V(t)$. Noise-free complex harmonic signals at $f_\textrm{g}$ and $f_\textrm{rf}-f_\textrm{g}$ can then be constructed for the demodulation of $n_\textrm{g}(t)$ and $n_\textrm{SB-}(t)$. As Eq. 3 is implemented numerically on full bandwidth signals, the effective demodulation bandwidth is 15 GHz. The low pass filter $\mathcal{F}_\textrm{low}$ used to reject the image frequencies is set at $f_g$, with is a good compromise between time-resolution and rejection capability. 

The imperfect frequency selectivity of $\mathcal{F}_\textrm{low}$ in Eq.~3 leads to some spectral leakage of $n_\textrm{rf}$ into the estimation of the population of the SB$^-$ mode. This leakage affects $n_\textrm{SB-}(t)$ only during the rise time $n_\textrm{rf}$, and leads to the artefact at $t\in [-3,~3] $ ns that can be perceived in the $n_\textrm{SB-}(t)$ curves of Fig.~\ref{Fig_TimeResolvedTransients}. Since $n_\textrm{rf}$ is reproducible, this leakage artefact is systematic but reproducible. It can be corrected for by systematically subtracting a low-passed fraction of $n_\textrm{rf}$ when estimating $n_\textrm{SB}(t)$. This correction is done in Fig. 4(b). To minimize the detrimental influence of this article on the determination of $\tau_{SB-}$ when it approaches the  $[-3,~3] $ ns interval, we have used a slightly larger threshold of 80\% (i.e. -2 dB) instead of 70\% (-3 dB) for the determination of $\tau_{SB-}$. This statistically increments our $\tau_{SB-}$ by circa 0.5 ns, a value that we consider as negligible compared to the errors arising from the experimental noise. 

\subsection*{Appendix D: Stack composition}
The device under investigation is a magnetic tunnel junction, composed of a magnetic multilayer stack 6 IrMn / 2 CoFe30 / 0.825 Ru / 2.6 CoFe40B20 / MgO [8 $\Omega. \mu\textrm m ^2$] / 2.0 CoFe40B20 / 0.21 Ta / 40 CoFeSiB / 10 Ta / 7 Ru (thickness in nm).

%\bibliography{bib.bib}
%apsrev4-2.bst 2019-01-14 (MD) hand-edited version of apsrev4-1.bst
%Control: key (0)
%Control: author (8) initials jnrlst
%Control: editor formatted (1) identically to author
%Control: production of article title (0) allowed
%Control: page (0) single
%Control: year (1) truncated
%Control: production of eprint (0) enabled
%

\end{document}